\documentclass[preprint,notoc]{JHEP3}
\usepackage{epsfig}

\def\etmiss{E_T^{\rm miss}}
\def\eslt{E_T^{\rm miss}}

\def\to{\rightarrow}

\def\bi{\begin{itemize}}
 \def\ei{\end{itemize}}

\def\c1p{C1^\prime}

\def\tg{\tilde g}

\def\tq{\tilde q}

\def\alt{\stackrel{<}{\sim}}
\def\agt{\stackrel{>}{\sim}}
\def\be{\begin{equation}}  
\def\ee{\end{equation}}  
\def\bea{\begin{eqnarray}}  
\def\eea{\end{eqnarray}}  

\def\sps1ap{SPS1a$^\prime$}
\title{LHC discovery potential for supersymmetry\\
with $\sqrt{s}=$7 TeV and 5-30 fb$^{-1}$
}
\author{Howard Baer$^{a}$, Vernon Barger$^b$, Andre Lessa$^c$ 
and Xerxes Tata$^{d}$\\
$^a$Dept.\ of Physics and Astronomy, University of Oklahoma, Norman, OK 73019, USA\\
$^b$Dep't of Physics, University of Wisconsin, Madison, WI 53706, USA\\
$^c$Instituto de F\'isica, Universidade de S\~ao Paulo, S\~ao Paulo - SP, Brazil\\
$^d$Dept. of Physics and Astronomy, University of Hawaii, Honolulu, HI 96822, US\\
E-mail: \email{baer@nhn.ou.edu}, \email{barger@pheno.wisc.edu},
\email{lessa@fma.if.usp.br}, \email{tata@phys.hawaii.edu}}

\preprint{\vbox{UH-511-1180-11}}

\abstract{ We extend our earlier results delineating the supersymmetry
  (SUSY) reach of the CERN Large Hadron Collider operating at a
  centre-of-mass energy $\sqrt{s}=7$ TeV to integrated luminosities in
  the range 5 - 30~fb$^{-1}$.  Our results are presented within the
  paradigm minimal supergravity model (mSUGRA or CMSSM).  Using a
  6-dimensional grid of cuts for the optimization of signal to
  background ratio -- including missing $E_T$-- we find for $m_{\tg}\sim
  m_{\tq}$ an LHC $5\sigma$ SUSY discovery reach of $m_{\tg}\sim 1.3,\
  1.4,\ 1.5$ and 1.6 TeV for 5, 10, 20 and 30 fb$^{-1}$, respectively.
  For $m_{\tq}\gg m_{\tg}$, the corresponding reach is instead
  $m_{\tg}\sim 0.8,\ 0.9,\ 1.0$ and 1.05 TeV, for the same integrated
  luminosities.} 


\begin{document}

\section{Introduction}
\label{sec:intro}

In 2011, the CERN Large Hadron Collider has produced proton-proton
collisions at a centre-of-mass energy $\sqrt{s}= 7$~TeV (LHC7) and has
enabled both ATLAS and CMS experiments to each accumulate over
5~fb$^{-1}$ of useful data. The current plan is to resume running with
$pp$ collisions in early 2012, with a goal to amass in the vicinity of
10-30 fb$^{-1}$ of usable data.  The 2012 run will likely be followed by
a shut down for $\sim 2.5$ years so that various upgrades may be
implemented; turn-on at or near design energy of $\sqrt{s}=14$ TeV is
then expected around 2015.

While many LHC analyses are focused on the elusive Higgs
boson, the search for weak scale supersymmetry (SUSY) \cite{wss}
remains an important part of the LHC program.
In a previous paper\cite{lhc7}, we presented projections for the LHC7
$5\sigma$ discovery reach for SUSY in the paradigm minimal supergravity
(mSUGRA or CMSSM) model\cite{msugra}.  In that study, we presented
discovery strategies for early SUSY discovery and made projections for
the LHC7 reach for a variety of integrated luminosities ranging from
100~pb$^{-1}$  up to 2~fb$^{-1}$, well beyond what was then expected
to be delivered in the entire 7~TeV run.
 LHC reach projections for $\sqrt{s}=14$ TeV (LHC14) have been reported
in earlier studies \cite{lhcreach}.

Recent analyses (performed within the mSUGRA model) of the LHC7 data
by the ATLAS\cite{atlas} and CMS\cite{cms} experiments based on just
$\sim 1$ fb$^{-1}$ of integrated luminosity have found no indication of
SUSY so far, yielding 95\% CL lower limits of $m_{\tq}\sim m_{\tg}\agt
1$ TeV for comparable gluino and squark masses, and $m_{\tg}\agt 0.6$
TeV for the case where $m_{\tq}\gg m_{\tg}$.  It is worth emphasizing
that although all squarks are by assumption degenerate within the mSUGRA
framework, the squark mass limit cited above arises mostly from signals
for first generation squarks that are much more copiously produced from
$qq$ and $qg$ initial states than their second and third generation
cousins. In other words, the LHC7 squark limit really applies to up and
down type squarks -- other squark flavors may be significantly lighter
than the quoted bounds. These LHC7 bounds {\it do not apply} to third
generation squarks or to electroweak-inos, the only sparticles with
significant couplings to the Higgs sector and to which the naturalness
arguments that yield upper mass bounds on sparticles apply. Indeed
models with ${\cal O}(10-100)$~TeV gluinos and first generation sfermions
but with sub-TeV third generation sfermions and electroweak-inos
\cite{esusy} that have been proposed to ameliorate the SUSY flavour and
$CP$ problems are not in conflict with these LHC7 data.

The LHC has performed spectacularly and has already delivered an
integrated luminosity of 5~fb$^{-1}$ and, as we mentioned, is expected to
deliver a comparable or larger data set in 2012. 
This motivated us to extend our
earlier projections\cite{lhc7}  of the LHC7 reach
for SUSY to integrated luminosities up to 30~fb$^{-1}$.
As before, we work within the mSUGRA framework, the 
parameter space of which is given by,
\be
m_0,\ m_{1/2},\ A_0,\ \tan\beta ,\ sign(\mu )\;.
\ee
Here, $m_0$ is a common GUT scale soft SUSY breaking (SSB) scalar mass,
$m_{1/2}$ is a common GUT scale soft supersymmetry breaking (SSB) gaugino mass, $A_0$ is a common GUT
scale trilinear SSB term, $\tan\beta$ is the ratio of Higgs field vevs,
and $\mu$ is the superpotential higgsino mass term, whose magnitude, but
not sign, is constrained by the electroweak symmetry breaking
minimization conditions.  

For each model parameter space point, many
simulated collider events are generated and compared against SM
backgrounds with the same experimental signature \cite{bg}. A 6-dimensional
grid of cuts\cite{lhc7} is then employed to enhance the SUSY signal over SM
backgrounds, and the signal is deemed observable if it satisfies
pre-selected criteria for observability.  Based on previous
studies \cite{lhcreach}, we include in our analysis the following channels:
\begin{itemize}
\item $jets +\eslt$ (no isolated leptons),
\item $1\ell +jets+\eslt$,
\item two opposite-sign isolated leptons (OS)$+jets+\eslt$,
\item two same-sign isolated leptons (SS)$+jets+\eslt$,
\item $3\ell +jets+\eslt$.
\end{itemize}

For the simulation of the background events, we use AlpGen\cite{alpgen} 
to compute the hard scattering events and Pythia \cite{pythia} for the
subsequent showering and hadronization.  For the final states containing
multiple jets (namely $Z(\to ll,\nu\nu) + jets$, $W(\to l\nu) + jets$,
$b\bar{b} + jets$, $t\bar{t} + jets$, $Z + b\bar{b} + jets$, $Z +
t\bar{t} + jets$, $W + b\bar{b} + jets$, $W + t\bar{t} + jets$ and QCD),
we use the MLM matching algorithm to avoid double counting.
All the processes included in our analysis are shown in
Table~1 of Ref.~\cite{lhc7} as well as their total cross-sections, number of
events generated and event generator used. Here, we show in Table \ref{tab:BG}
the various backgrounds along with $k$-factors\footnote{By $k$-factor, here
we actually mean $\sigma^{NLO}/\sigma^{LO}$. Normally, one compares the two
cross sections using an identical renormalization/factorization scale for the
two cases. Here, we merely compute $\sigma^{LO}$
using AlpGen and $\sigma^{NLO}$
using MCFM, using the pre-programmed default scale choices for the latter.
} 
used to normalize the
generator cross sections to NLO QCD results where available.  
The background $k$-factors were computed using MCFM\cite{mcfm} for the NLO
cross sections, and AlpGen for the LO ones.

The signal events were generated using Isajet 7.79\cite{isajet} which,
given a mSUGRA parameter set, generates all $2\to 2$ SUSY processes in
the right proportion, and decays the sparticles to lighter sparticles
using the appropriate branching ratios
and decay matrix elements, until the parent sparticle cascade decay\cite{cascade} 
terminates in the stable lightest supersymmetric
particle (LSP), assumed here to be the lightest neutralino.  
Total gluino and squark production cross
sections have been presented in Ref. \cite{lhc7} at NLO QCD using
Prospino\cite{prospino}, and will not be repeated here.  It is worth
noting that for $m_{\tq}\sim m_{\tg}$, $\tg\tq$ associated production is
the dominant strongly interacting SUSY production mechanism, while for
$m_{\tq}\gg m_{\tg}$, $\tg\tg$ pair production tends to dominate.

For event generation, we use a toy detector simulation with calorimeter
cell size $\Delta\eta\times\Delta\phi=0.05\times 0.05$ and $-5<\eta<5$
. The HCAL (hadronic calorimetry) energy resolution is taken to be
$80\%/\sqrt{E}\oplus 3\%$ for $|\eta|<2.6$ and FCAL (forward
calorimetry) is $100\%/\sqrt{E}\oplus 5\%$ for $|\eta|>2.6$, where
$\oplus$ denotes a combination in quadrature. The ECAL (electromagnetic
calorimetry) energy resolution is assumed to be $3\%/\sqrt{E}\oplus 0.5\%$. We
use the cone-type Isajet \cite{isajet} jet-finding algorithm to group
the hadronic final states into jets. Jets and isolated lepton are
defined as follows: 

\bi
\item Jets are hadronic clusters with $|\eta| < 3.0$,
$R\equiv\sqrt{\Delta\eta^2+\Delta\phi^2}\leq0.4$ and $E_T(jet)>50$ GeV.
\item Electrons and muons are considered isolated if they have $|\eta| <
2.0$, $p_T(l)>10 $ GeV with visible activity within a cone of $\Delta
R<0.2$ about the lepton direction, $\Sigma E_T^{cells}<5$ GeV.  
\item  We identify hadronic clusters as 
$b$-jets if they contain a B hadron with $E_T(B)>$ 15 GeV, $\eta(B)<$ 3 and
$\Delta R(B,jet)<$ 0.5. We assume a tagging efficiency of 60$\%$ and 
light quark and gluon jets can be mis-tagged
as a $b$-jet with a probability 1/150 for $E_{T} \leq$ 100 GeV,
1/50 for $E_{T} \geq$ 250 GeV, with a linear interpolation
for 100 GeV $\leq E_{T} \leq$ 250 GeV \cite{btag}.
\ei

\begin{table}
\centering
\begin{tabular}{|l|c|}
\hline
SM process & $k$-factor \\
\hline
$t\bar{t}$ & 0.99 \\
$Z/\gamma$ + jets & 1.47 \\
$W$ + jets & 1.53 \\
$Z(\to \nu\bar{\nu})$ + $b\bar{b}$ & 1.18 \\
$Z/\gamma (\to l\bar{l})$ + $b\bar{b}$ & 1.03 \\
$WW$ & 1.38 \\
$WZ$ & 1.47 \\
$ZZ$ & 1.35\\
\hline
\end{tabular}
\begin{tabular}{|l|c|}
\hline
SM process & $k$-factor \\
\hline
$QCD, b\bar{b}$ & -- \\
 $Z$ + $t\bar{t}$ & -- \\
 $W$ + $t\bar{t}$ & -- \\
 $W$ + $b\bar{b}$ & -- \\
 $W$ + $tb$ & -- \\
$t\bar{t}t\bar{t}$ & -- \\
$t\bar{t}b\bar{b}$  & -- \\
$b\bar{b}b\bar{b}$ & -- \\
\hline
\end{tabular}
\caption{Background processes included in this LHC7 study, along with
the $k$-factor (from MCFM and AlpGen) used (when available) to normalize
to NLO QCD.  For $t\bar{t}$ production, the renormalization scale is
chosen to match the NLO cross section. The event generator used, total
cross sections and number of generated events are listed in Table 1 of
Ref. \cite{lhc7}. All light (and {\it b}) partons in the final state are
required to have $E_T> 40$~GeV. For QCD, we generate the hardest final
parton jet in distinct bins to get a better statistical representation
of hard events.}
\label{tab:BG}
\end{table}

As in Ref.~\cite{lhc7}, we
define the signal to be observable if
\begin{description}
  \item[] \qquad \qquad $S \ge max\left[5\sqrt{B},\ 5,\ 0.2B\right]$
\end{description}
where $S$ and $B$ are the expected number of signal and background
events, respectively, for an assumed value of integrated luminosity. 
The requirement $S\ge 0.2B$ is imposed to avoid
the possibility that a {\it small} signal on top of a {\it large} background
could otherwise be regarded as
statistically significant, but whose viability would require
the background level to be known with 
exquisite precision in order to establish a discovery. 
For cases with very low signal and BG event numbers, we require the 
Poisson probability to correspond to the $5\sigma$ level. 
%

The grid of cuts used in our optimized analysis is:
\bi
  \item $\etmiss >$ 50,100 - 1000 GeV (in steps of 100 GeV),
  \item $n(jets) \geq$ 2, 3, 4, 5 or 6,
  \item $n(b-jets) \geq$ 0, 1, 2 or 3,
  \item $E_T(j_1) >$ 50 - 300 GeV (in steps of 50 GeV) and 400-1000 GeV 
(in steps of 100 GeV) (jets are ordered $j_1-j_n$, from highest to lowest $E_T$),
  \item $E_T(j_2) >$ 50 - 200 GeV (in steps of 30 GeV) and 300, 400, 500 GeV,
  \item $n(\ell ) = $ 0, 1, 2, 3, OS, SS and inclusive channel: $n(\ell) \geq$ 0.
(Here, $\ell = e,\ \mu$). 
  \item 10 GeV$\le m(\ell^+\ell^-) \le 75$ GeV or $m(\ell^+\ell^-) \ge 105$ GeV 
(for the OS, same flavor (SF) dileptons only),
  \item transverse sphericity $S_T > 0.2$.
\ei

\FIGURE[tbh]{
\includegraphics[width=13cm,clip]{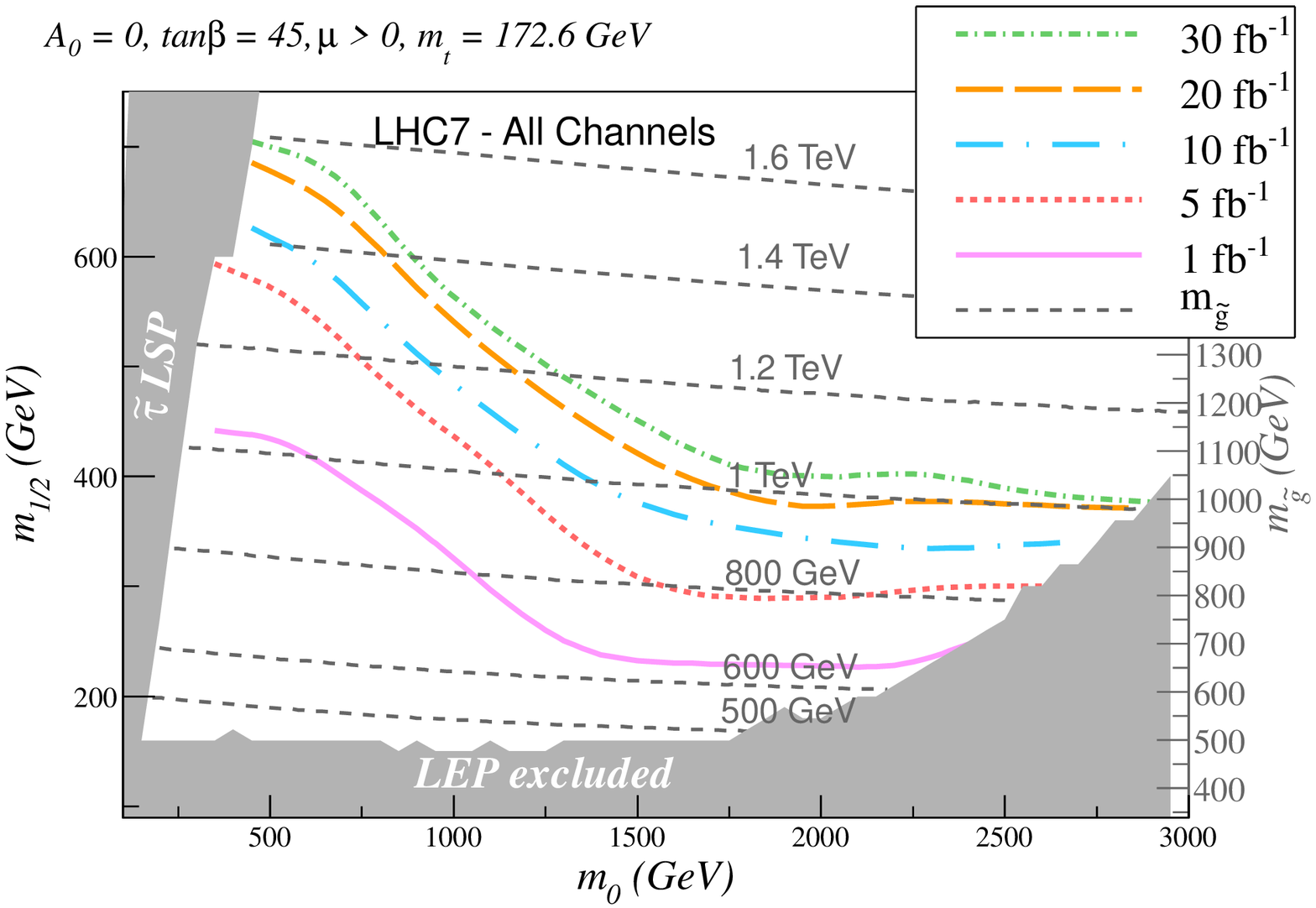}
\caption{ The optimized SUSY reach of LHC7 for different integrated
luminosities combining the different channels described in the text. The
signal is observable if it falls below the solid contour for the
corresponding integrated luminosity.
The fixed mSUGRA parameters are $A_0=0$, $\tan\beta =45$ and $\mu >0$.
Gluino mass contours are shown by the dashed, dark
grey curves.  
The shaded grey area is excluded due to
stau LSPs (left side of figure) or no electroweak symmetry breaking
(right side of figure), while the shaded grey area marked ``LEP excluded''
is excluded by non-observation of a sparticle signal from LEP2
searches.
All sparticle and background cross sections are normalized to NLO QCD
values via $k$-factors.
}\label{fig:reach}}

We show in Fig.~\ref{fig:reach} the optimized $5\sigma$ discovery reach
of LHC7 for various choices of integrated luminosity in the $m_0\ vs.\
m_{1/2}$ plane.  We also take $A_0=0$, $\tan\beta =45$ and $\mu >0$,
with $m_t=172.6$ GeV.\footnote{Recent evidence from Atlas\cite{atlas_h}
and CMS\cite{cms_h} using 5 fb$^{-1}$ of data show some evidence for a
Higgs scalar $h$ with $m_h\sim 125$ GeV. For $A_0=0$, it is very difficult 
to accommodate such a Higgs mass in the mSUGRA model. For $A_0\sim \pm 2m_0$,
then $m_h\sim 125$ GeV can be accommodated, but mainly at rather high $m_0\sim 2-10$ TeV.
For more details, see {\it e.g.} Ref.~\cite{h125}. Our reach projections are largely
insensitive to variation in $A_0$ (and subsequent small changes in $m_h$) as explained below.
}
Gluino iso-mass contours are shown, as obtained
using the ISASUGRA routines\cite{isasugra} in Isajet. We see from
Fig.~\ref{fig:reach} that with $\sim 1$~fb$^{-1}$ of integrated
luminosity, the LHC7 sensitivity does indeed extend to $m_{\tg}\sim 1.1$
TeV for $m_{\tq}\sim m_{\tg}$, and to $m_{\tg}\sim 0.65$ TeV for
$m_{\tq}\gg m_{\tg}$\footnote{We stress that the curves presented here
include an optimization over several search channels and correspond to a
$5\sigma$ discovery reach. Care must be taken when comparing these
results with experimental bounds, which are usually presented for single
channels at $95\%$ C.L. ($\sim 2\sigma$).}.  For 5 fb$^{-1}$ of
integrated luminosity (for which we expect ATLAS and CMS analyses in
Spring 2012), the LHC discovery reach extends to $m_{\tg}\sim 1.3$ TeV
for $m_{\tq}\sim m_{\tg}$, and to $m_{\tg}\sim 0.8$ TeV for $m_{\tq}\gg
m_{\tg}$.  As integrated luminosity moves into the 20-30 fb$^{-1}$
regime, the LHC7 reach for $m_{\tq}\sim m_{\tg}$ moves up to
$m_{\tg}\sim 1.5-1.6$ TeV. For the case where $m_{\tq}\gg m_{\tg}$, the
20-30 fb$^{-1}$ LHC reach approaches $m_{\tg}\sim 1$ TeV.
We stress that -- as
discussed above -- while non-observation of the signal at LHC7 may
qualitatively point toward very heavy gluinos and first generation squarks,
{\it this does not} in and of itself preclude SUSY as the new physics
that stabilizes the weak scale \cite{esusy} because third generation
squarks and electroweak-inos could still be at the sub-TeV scale.


While our results are presented for the particular choice of mSUGRA
parameters $A_0=0$ and $\tan\beta =45$, we emphasize here that we expect
these results to hold also for other choices of $A_0$ and $\tan\beta$,
and also for $\mu <0$.  Variation of $A_0$ mainly affects third
generation sparticle masses, while the reach is  determined
mostly by $m_{\tg}$ and the first generation squark masses.  
Moreover, variation of $\tan\beta$ mainly
affects the size of $b$ and $\tau$ Yukawa couplings, and these feed only
weakly into the reach plots: for instance, sparticle decays to third
generation matter are enhanced at large $\tan\beta$\cite{ltanb} where
$b$-tagging may somewhat enhance the LHC reach for gluinos \cite{enhb}
as already demonstrated by ATLAS \cite{atlasb}.

To give the reader an idea of the dominant event topologies in which
experiments at LHC7 will be able to probe SUSY in the 2012 run, we show
in Fig.~\ref{fig:topo} the optimized $5\sigma$ 
reach via the $0\ell$, $1\ell$, OS
dilepton, SS dilepton and the trilepton channels for 20~fb$^{-1}$.  The
striking feature of the figure is that while the reach is dominated by
the low multiplicity ($n_{\ell}=0, 1$) lepton channels for $m_0\alt
1.5$~TeV, the reach in the low background but rate-limited trilepton
channel becomes competitive with that in other channels if squarks are
essentally decoupled at LHC7 as could well be the case. We have checked
that this is true also for an integrated luminosity of 10~fb$^{-1}$.  

\FIGURE[tbh]{
\includegraphics[width=13cm,clip]{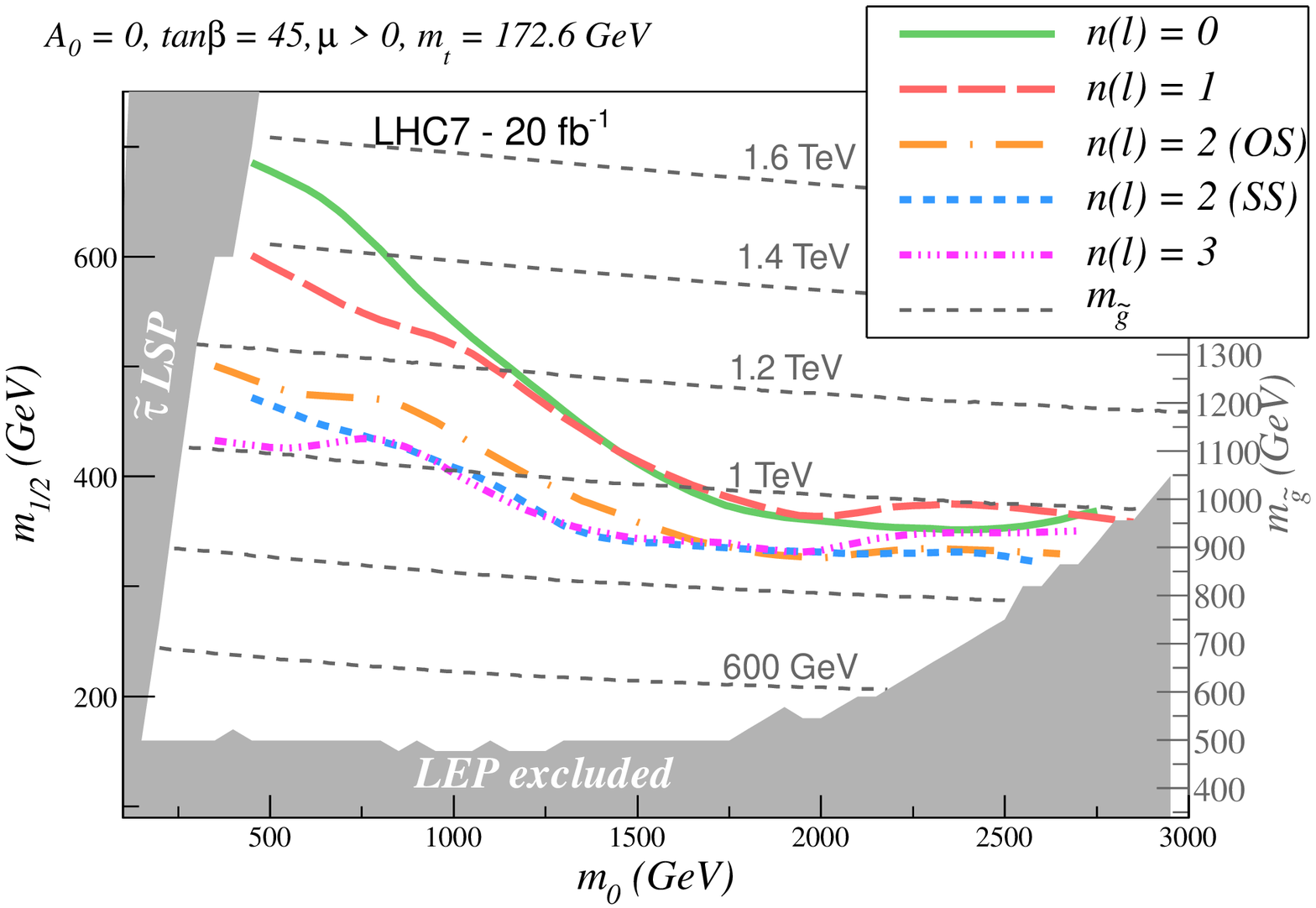}
\caption{ The optimized $5\sigma$ SUSY reach of LHC7 in various channels
  classified by lepton multiplicity: $0\ell$, 1$\ell$, $SS$ dilepton,
  $OS$ dilepton and trilepton for an integrated luminosity of
  20~fb$^{-1}$. 
Any mSUGRA point will be
  observable if it falls below the corresponding contour.
The fixed mSUGRA parameters are $A_0=0$, $\tan\beta =45$ and $\mu >0$.
Gluino mass contours are shown by the dashed, dark
grey curves.  
The shaded grey area is excluded due to
stau LSPs (left side of figure) or no electroweak symmetry breaking
(right side of figure), while the shaded grey area marked ``LEP excluded''
is excluded by non-observation of a sparticle signal from LEP2
searches.
All sparticle and background cross sections are normalized to NLO QCD
values via $k$-factors.
}\label{fig:topo}}

In summary, we have presented updated $5\sigma$ discovery contours for
the paradigm mSUGRA/CMSSM SUSY model for LHC7 with 5-30 fb$^{-1}$ of
integrated luminosity.  These results help to understand the
capabilities of LHC7 for discovering supersymmetry in 2012-2013. Within
mSUGRA, for integrated luminosity 20-30 fb$^{-1}$, we expect LHC7 to
probe $m_{\tg}$ up to $\sim 1.6$ TeV for $m_{\tq}\simeq m_{\tg}$, while
we expect LHC7 to probe up to $m_{\tg}\sim 1$ TeV for $m_{\tq}\gg
m_{\tg}$. If squarks are much heavier than gluinos, the reach at LHC7
via the inclusive trilepton channel will be competitive in reach with
the canonical jets plus $\eslt$ channel.

\acknowledgments

This work was supported by the United States Department of Energy
and by Fundac\~ao de Apoio \`a Pesquisa do Estado de S\~ao Paulo (FAPESP).


%

\end{document}